\documentclass[twocolumn,aps,unsortedaddress,superscriptaddress,prl,floatfix,showpacs]{revtex4}
\usepackage{epsfig}

\begin{document}


\title{Measurement of $\Upsilon$ Production
for $p+p$ and $p+d$ Interactions at 800 GeV/c}

\affiliation{Abilene Christian University, Abilene, TX 79699}
\affiliation{Physics Division, Argonne National Laboratory, Argonne, IL 60439}
\affiliation{Fermi National Accelerator Laboratory, Batavia, IL 60510}
\affiliation{Georgia State University, Atlanta, GA 30303}
\affiliation{Illinois Institute of Technology, Chicago, IL  60616}
\affiliation{University of Illinois at Urbana-Champaign, Urbana, IL 61801}
\affiliation{Los Alamos National Laboratory, Los Alamos, NM 87545}
\affiliation{University of New Mexico, Albuquerque, NM 87131}
\affiliation{New Mexico State University, Las Cruces, NM 88003}
\affiliation{Oak Ridge National Laboratory, Oak Ridge, TN 37831}
\affiliation{Texas A\&M University, College Station, TX 77843}
\affiliation{Valparaiso University, Valparaiso, IN 46383}

\author{L.Y.~Zhu}
\affiliation{University of Illinois at Urbana-Champaign, Urbana, IL 61801}

\author{P.E.~Reimer}
\affiliation{Physics Division, Argonne National Laboratory, Argonne, IL 60439}
\affiliation{Los Alamos National Laboratory, Los Alamos, NM 87545}

\author{B.A.~Mueller}
\affiliation{Physics Division, Argonne National Laboratory, Argonne, IL 60439}

\author{T.C.~Awes}
\affiliation{Oak Ridge National Laboratory, Oak Ridge, TN 37831}

\author{M.L.~Brooks}
\affiliation{Los Alamos National Laboratory, Los Alamos, NM 87545}

\author{C.N.~Brown}
\affiliation{Fermi National Accelerator Laboratory, Batavia, IL 60510}

\author{J.D.~Bush}
\affiliation{Abilene Christian University, Abilene, TX 79699}

\author{T.A.~Carey}
\affiliation{Los Alamos National Laboratory, Los Alamos, NM 87545}

\author{T.H.~Chang}
\affiliation{New Mexico State University, Las Cruces, NM 88003}

\author{W.E.~Cooper}
\affiliation{Fermi National Accelerator Laboratory, Batavia, IL 60510}

\author{C.A.~Gagliardi}
\affiliation{Texas A\&M University, College Station, TX 77843}

\author{G.T.~Garvey}
\affiliation{Los Alamos National Laboratory, Los Alamos, NM 87545}

\author{D.F.~Geesaman}
\affiliation{Physics Division, Argonne National Laboratory, Argonne, IL 60439}

\author{E.A.~Hawker}
\affiliation{Texas A\&M University, College Station, TX 77843}

\author{X.C.~He}
\affiliation{Georgia State University, Atlanta, GA 30303}

\author{D.E.~Howell}
\affiliation{University of Illinois at Urbana-Champaign, Urbana, IL 61801}

\author{L.D.~Isenhower}
\affiliation{Abilene Christian University, Abilene, TX 79699}

\author{D.M.~Kaplan}
\affiliation{Illinois Institute of Technology, Chicago, IL  60616}

\author{S.B.~Kaufman}
\affiliation{Physics Division, Argonne National Laboratory, Argonne, IL 60439}

\author{S.A.~Klinksiek}
\affiliation{University of New Mexico, Albuquerque, NM 87131}

\author{D.D.~Koetke}
\affiliation{Valparaiso University, Valparaiso, IN 46383}

\author{D.M.~Lee}
\affiliation{Los Alamos National Laboratory, Los Alamos, NM 87545}

\author{W.M.~Lee}
\affiliation{Fermi National Accelerator Laboratory, Batavia, IL 60510}
\affiliation{Georgia State University, Atlanta, GA 30303}

\author{M.J.~Leitch}
\affiliation{Los Alamos National Laboratory, Los Alamos, NM 87545}

\author{N.~Makins}
\affiliation{Physics Division, Argonne National Laboratory, Argonne, IL 60439}
\affiliation{University of Illinois at Urbana-Champaign, Urbana, IL 61801}

\author{P.L.~McGaughey}
\affiliation{Los Alamos National Laboratory, Los Alamos, NM 87545}

\author{J.M.~Moss}
\affiliation{Los Alamos National Laboratory, Los Alamos, NM 87545}

\author{P.M.~Nord}
\affiliation{Valparaiso University, Valparaiso, IN 46383}

\author{V.~Papavassiliou}
\affiliation{New Mexico State University, Las Cruces, NM 88003}

\author{B.K.~Park}
\affiliation{Los Alamos National Laboratory, Los Alamos, NM 87545}

\author{G.~Petitt}
\affiliation{Georgia State University, Atlanta, GA 30303}

\author{J.C.~Peng}
\affiliation{University of Illinois at Urbana-Champaign, Urbana, IL 61801}
\affiliation{Los Alamos National Laboratory, Los Alamos, NM 87545}

\author{M.E.~Sadler}
\affiliation{Abilene Christian University, Abilene, TX 79699}

\author{W.E.~Sondheim}
\affiliation{Los Alamos National Laboratory, Los Alamos, NM 87545}

\author{P.W.~Stankus}
\affiliation{Oak Ridge National Laboratory, Oak Ridge, TN 37831}

\author{T.N.~Thompson}
\affiliation{Los Alamos National Laboratory, Los Alamos, NM 87545}

\author{R.S.~Towell}
\affiliation{Abilene Christian University, Abilene, TX 79699}

\author{R.E.~Tribble}
\affiliation{Texas A\&M University, College Station, TX 77843}

\author{M.A.~Vasiliev}
\affiliation{Texas A\&M University, College Station, TX 77843}

\author{J.C.~Webb}
\affiliation{New Mexico State University, Las Cruces, NM 88003}

\author{J.L.~Willis}
\affiliation{Abilene Christian University, Abilene, TX 79699}

\author{P.~Winter}
\affiliation{Physics Division, Argonne National Laboratory, Argonne, IL 60439}

\author{D.K.~Wise}
\affiliation{Abilene Christian University, Abilene, TX 79699}

\author{Y.~Yin}
\affiliation{University of Illinois at Urbana-Champaign, Urbana, IL 61801}

\author{G.R.~Young}
\affiliation{Oak Ridge National Laboratory, Oak Ridge, TN 37831}

\collaboration{FNAL E866/NuSea Collaboration}
\noaffiliation

\date{\today}

\begin{abstract}

We report a high statistics measurement of $\Upsilon$ production with an 800
GeV/c proton beam on hydrogen and deuterium targets. The dominance of the
gluon-gluon fusion process for $\Upsilon$ production at this energy implies 
that the cross section ratio, $\sigma (p + d \to \Upsilon) / 2\sigma 
(p + p\to \Upsilon)$, is sensitive to the gluon content in the neutron relative 
to that in the proton. Over the kinematic region $0 < x_F < 0.6$, this 
ratio is found to be 
consistent with unity, in striking contrast to the behavior
of the Drell-Yan cross section ratio $\sigma(p+d)_{DY}/2\sigma(p+p)_{DY}$. 
This result shows that the gluon distributions in the proton and neutron 
are very similar. The $\Upsilon$ production cross sections are also 
compared with the $p+d$ and $p+{\rm Cu}$ cross sections from earlier 
measurements.

\end{abstract} 

\maketitle

In the CERN NA51~\cite{na51a} and Fermilab E866/NuSea~\cite{hawker,peng,towell}
experiments on proton-induced dimuon production, a striking difference was
observed for the Drell-Yan cross sections between $p+p$ and $p+d$.
As the underlying mechanism for the Drell-Yan process involves
quark-antiquark annihilation, this difference has been attributed to
the asymmetry between the up and down sea quark distributions
in the proton. From the $\sigma (p+d)_{DY}/2\sigma(p+p)_{DY}$ ratios 
the Bjorken-$x$ dependence of the sea-quark
$\bar d / \bar u$ flavor
asymmetry has been extracted~\cite{na51a,hawker,peng,towell}.

The Fermilab E866 dimuon experiment also recorded a large number of
$\Upsilon \to \mu^+ \mu^-$ events. In this paper, we present 
results on the $\Upsilon$
differential cross sections for $p+p$ and $p+d$ 
over the kinematic range $0<p_T<3.5$ GeV/c
and $0 < x_F < 0.6$. Unlike the electromagnetic
Drell-Yan process, quarkonium
production is a strong interaction dominated by the subprocess of
gluon-gluon fusion at this beam energy~\cite{jansen,vogt99}. Therefore, the
quarkonium production cross
sections are primarily sensitive to the gluon distributions in the
colliding hadrons. The $\Upsilon$ production ratio, $\sigma(p+d \to \Upsilon)
/2 \sigma(p+p \to \Upsilon)$, is expected to
probe the gluon content in the neutron relative to that in the 
proton~\cite{piller}. 
As pointed out by Piller and Thomas~\cite{piller}, charge symmetry
violation at the parton level could lead to a different gluon distribution 
in the proton versus that in the neutron. A precise measurement of the
$\sigma(p+d \to \Upsilon)/2 \sigma(p+p \to \Upsilon)$ ratios would provide
a constraint on the effect of charge symmetry violation on the gluon
distributions.

High statistics $\Upsilon$ production cross sections at 800 GeV have been
reported for $p+d$~\cite{pat} and $p+{\rm Cu}$~\cite{e605}. The per-nucleon 
cross sections for $p+d$ were roughly a factor of two greater than those 
of $p+{\rm Cu}$. Such a difference could not be explained by the nuclear
effect of $\Upsilon$ production, which was found to be small~\cite{e772ups}.
The $p+p$ and $p+d$ data would shed new light on this appararent discrepancy.
Moreover, $\Upsilon$ production in the simple
$p+p$ and $p+d$ systems provides the baseline information for future
searches of possible suppression of $\Upsilon$ production as a signature for
a quark-gluon plasma in relativistic heavy-ion 
collisions~\cite{karsch,bedjidian}.

The experiment was performed at Fermilab using the upgraded Meson-East
magnetic pair spectrometer.
Details of the experimental setup can be
found elsewhere~\cite{towell,brown,towell1}. A primary proton beam with up to
$2 \times 10^{12}$ protons per 20-second beam spill was incident upon
one of three identical 50.8-cm long cylindrical stainless steel
target flasks containing either liquid hydrogen, liquid deuterium,
or vacuum. A copper beam dump located inside the second dipole
magnet absorbed the protons that passed through the target. Downstream
of the beam dump was a 13.4 interaction-length absorber wall of copper,
carbon and polyethylene that completely filled the aperture of the magnet.
This absorber wall removed hadrons produced in the target and the beam dump.

The targets alternated between hydrogen and deuterium every five beam
spills with a single spill collected on the empty flask at each target
change. Beam intensity was monitored by secondary-emission detectors, an
ion chamber, and quarter-wave RF cavities. Two scintillator telescopes
viewing the targets at $90^\circ$ monitored the luminosity, beam duty factor
and data acquisition live-time. The detector system consisted of
four tracking stations and a momentum analyzing magnet. The
trigger required a pair of triple hodoscope coincidences having a pattern
consistent with a muon pair from the target. 

\begin{figure}[tb]
\includegraphics*[width=\linewidth]{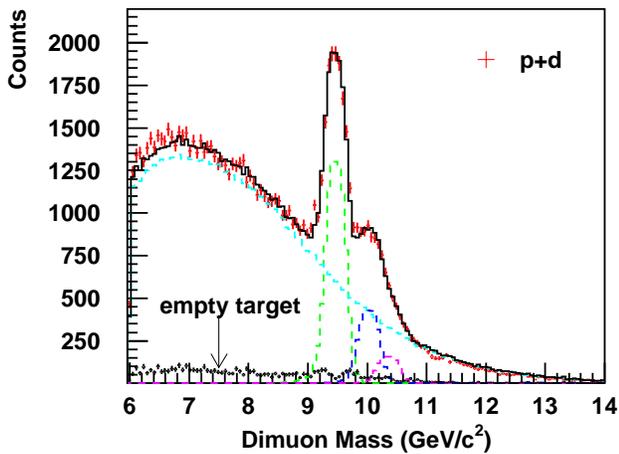}
\caption{(color online). Dimuon mass spectrum for $p+d$ interactions at
800 GeV/c. Fit to the spectrum is shown as the solid curve. Contributions
from the Drell-Yan continuum and $\Upsilon$ resonances are shown as dashed curves.
The luminosity weighted spectrum from the empty-target measurements is
also shown.}
\label{crosfig1}
\end{figure}

Tracks reconstructed in the drift chambers were extrapolated to the target
using the momentum determined by the analyzing magnet. The target
position was used to refine the parameters of each muon track. The resulting
rms mass resolution for the $\Upsilon$ resonances is $\approx$ 250 MeV. Monte
Carlo studies show that this resolution is dominated by the finite target
length and the multiple scattering of muons in the
absorber. Figure 1 shows the dimuon mass spectra for the high-mass
data collected with the deuterium target. The high-mass
data set contains approximately 20,000 $\Upsilon$ events.

\begin{table}[hbtp]
\caption{Product of the $\Upsilon$ production cross section per nucleon 
and the $\Upsilon \rightarrow \mu^+ \mu^- $ branching ratio ($B$) for $p+d$ 
and $p+p$ interactions at 800 GeV/c. The uncertainties are statistical 
only. The values of $p_0$ are also listed.}
\label{result}
\begin{center}
\begin{tabular}{|c|ccc|}
\hline \hline
$x_F$ & \multicolumn{3}{c|}{$B \cdot d\sigma/dx_F({\rm pb})$ for 
$p d \rightarrow \mu^+ \mu^- X $}  \\
\cline{2-4}
 &  $\Upsilon(1S)$ & $\Upsilon(2S)$ & $\Upsilon(3S)$   \\
\hline
 0.05 & 3.246$\pm$0.119 & 0.969$\pm$0.081 & 0.529$\pm$0.064 \\
 0.15 & 2.963$\pm$0.080 & 0.863$\pm$0.054 & 0.250$\pm$0.042 \\
 0.25 & 1.934$\pm$0.059 & 0.666$\pm$0.041 & 0.224$\pm$0.031 \\
 0.35 & 1.253$\pm$0.043 & 0.453$\pm$0.032 & 0.177$\pm$0.026 \\
 0.45 & 0.620$\pm$0.030 & 0.240$\pm$0.024 & 0.075$\pm$0.020 \\
 0.55 & 0.227$\pm$0.021 & 0.095$\pm$0.018 & 0.046$\pm$0.014 \\
\hline
$p_T$(GeV/c) & \multicolumn{3}{c|}{$B \cdot d\sigma/dp_T^2({\rm pb/GeV^2/c^2})$ 
for $p d \rightarrow \mu^+ \mu
^- X $} \\
 \cline{2-4}
 &  $\Upsilon(1S)$ & $\Upsilon(2S)$ & $\Upsilon(3S)$   \\
 \hline
 0.25 & 0.8648$\pm$0.0381 & 0.3921$\pm$0.0307 & 0.1063$\pm$0.0246 \\
 0.75 & 0.6347$\pm$0.0184 & 0.2264$\pm$0.0143 & 0.0804$\pm$0.0116 \\
 1.25 & 0.3996$\pm$0.0112 & 0.1225$\pm$0.0083 & 0.0563$\pm$0.0064 \\
 1.75 & 0.1968$\pm$0.0070 & 0.0610$\pm$0.0049 & 0.0228$\pm$0.0037 \\
 2.25 & 0.0964$\pm$0.0043 & 0.0281$\pm$0.0029 & 0.0114$\pm$0.0022 \\
 2.75 & 0.0416$\pm$0.0027 & 0.0111$\pm$0.0017 & 0.0050$\pm$0.0014 \\
 3.25 & 0.0191$\pm$0.0017 & 0.0029$\pm$0.0011 & 0.0030$\pm$0.0009 \\
$p_0$(GeV/c) &  3.39$\pm$0.04 & 3.06$\pm$0.07 & 3.36$\pm$0.17 \\
\hline \hline
 $x_F$ & \multicolumn{3}{c|}{$B \cdot d\sigma/dx_F({\rm pb})$ for 
 $p p \rightarrow \mu^+ \mu^- X $}
 \\
  \cline{2-4}
 &  $\Upsilon(1S)$ & $\Upsilon(2S)$ & $\Upsilon(3S)$   \\
\hline
 0.05 & 3.435$\pm$0.182 & 0.946$\pm$0.119 & 0.454$\pm$0.092 \\
 0.15 & 3.025$\pm$0.119 & 0.731$\pm$0.079 & 0.408$\pm$0.065 \\
 0.25 & 1.946$\pm$0.086 & 0.601$\pm$0.061 & 0.292$\pm$0.051 \\
 0.35 & 1.397$\pm$0.065 & 0.334$\pm$0.046 & 0.181$\pm$0.037 \\
 0.45 & 0.652$\pm$0.046 & 0.214$\pm$0.035 & 0.066$\pm$0.028 \\
 0.55 & 0.253$\pm$0.031 & 0.098$\pm$0.024 & 0.038$\pm$0.019 \\
\hline
$p_T$(GeV/c) & \multicolumn{3}{c|}{$B \cdot d\sigma/dp_T^2({\rm pb/GeV^2/c^2})$ 
for $p p \rightarrow \mu^+ \mu
^- X $} \\
  \cline{2-4}
 &  $\Upsilon(1S)$ & $\Upsilon(2S)$ & $\Upsilon(3S)$   \\
\hline
 0.25 & 0.8754$\pm$0.0576 & 0.3381$\pm$0.0448 & 0.1102$\pm$0.0374 \\
 0.75 & 0.6482$\pm$0.0276 & 0.2057$\pm$0.0218 & 0.0662$\pm$0.0164 \\
 1.25 & 0.3934$\pm$0.0161 & 0.0972$\pm$0.0119 & 0.0594$\pm$0.0098 \\
 1.75 & 0.2167$\pm$0.0107 & 0.0467$\pm$0.0069 & 0.0334$\pm$0.0058 \\
 2.25 & 0.1008$\pm$0.0064 & 0.0236$\pm$0.0042 & 0.0188$\pm$0.0035 \\
 2.75 & 0.0437$\pm$0.0040 & 0.0146$\pm$0.0027 & 0.0013$\pm$0.0012 \\
 3.25 & 0.0232$\pm$0.0027 & 0.0063$\pm$0.0019 & 0.0035$\pm$0.0021 \\
$p_0$(GeV/c) &  3.47$\pm$0.05 & 3.20$\pm$0.15 & 3.22$\pm$0.18 \\
 \hline \hline
\end{tabular}
\end{center}
\end{table}
   
To extract the yields of the $\Upsilon$ resonances, the contributions
of the Drell-Yan continuum under the $\Upsilon$ resonances need to be
determined and subtracted. Monte Carlo simulations for Drell-Yan events
using next-to-leading order calculations and the MRS98~\cite{mrs98}
parton distributions, which reproduce the $\bar d / \bar u$
asymmetry observed in the E866 Drell-Yan data~\cite{towell,towell1},
were carried out. The line shapes of the three $\Upsilon$ resonances
were also calculated using Monte Carlo. For each
$x_F$ and $p_T$ bin, the normalization factors for the Drell-Yan and
the $\Upsilon$ resonances were adjusted
to fit the data. Contributions of dimuon events from the stainless steel
target flask were subtracted using data obtained with the empty-target
measurements. The Drell-Yan normalization
factors were found to be consistent with unity, showing good agreements
between the data and the Monte Carlo simulation.
The dimuon mass spectra are well described by the sum of the various
contributions considered in the analysis, as illustrated in Fig. 1 for
the $p+d$ data. The mass spectra for various $x_F$
and $p_T$ bins are also well described using this fitting procedure.

\begin{figure}[tb]
\includegraphics*[width=\linewidth]{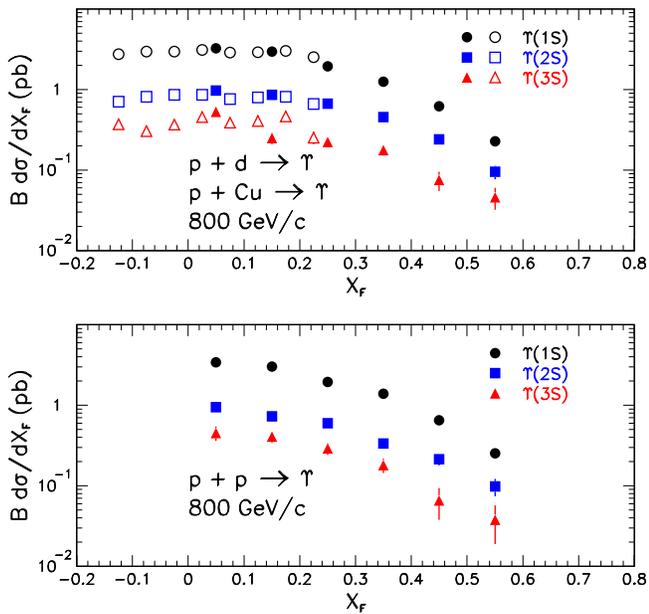}
\caption{(color online). Upper panel: $B\cdot d\sigma/x_F$ (per target nucleon)
for $\Upsilon(1S)$, $\Upsilon(2S)$,
and $\Upsilon(3S)$ production cross sections for $p+d$ at 800 GeV/c.
The E605 data~\cite{e605} for $p~+$ Cu are also shown as
open symbols. Lower panel:
$B\cdot d\sigma/x_F$ for $\Upsilon(1S)$, $\Upsilon(2S)$,
and $\Upsilon(3S)$ production for $p+p$ at 800 GeV/c.}
\label{crosfig2}
\end{figure}

The values of $B d\sigma/ d x_F$ per target nucleon for the three $\Upsilon$
resonances in $p+p$ and $p+d$ collisions are shown in Fig.\  2 and listed
in Tab.\  I ($B$ is the branching ratio for $\Upsilon
\to \mu^+ \mu^-$ decay). The $d\sigma/ d x_F$ differential cross sections are
obtained by integrating over $p_T$ using a $p_T$ distribution which
best fits the data. A $\pm 6.5$\% overall normalization
uncertainty, common to both the $p+p$ and $p+d$ cross sections, is
associated with the determination of the beam intensity~\cite{towell}.
Other systematic errors due to the uncertainty of the magnetic fields
of the spectrometer and the hodoscope efficiency are estimated to be
$\pm $3\%. Existing E605 data~\cite{e605}
for $p~+$ Cu collision covering the kinematic 
range $-0.15 < x_F < 0.25$
are also shown for comparison. The good agreement between
the E866 $p+d$ and the E605 
$p~+$ Cu data is consistent with the $A$-dependence
measurement performed by E772~\cite{e772ups}, which showed 
that the cross section is proportional to $A^\alpha$ with 
$\alpha \approx 0.962$.
Figure 2 also shows
that the relative yields for producing the $\Upsilon (1S)$, $\Upsilon (2S)$,
and $\Upsilon (3S)$ resonances are very similar for $p+d$ and $p~+$ Cu,
consistent with no significant nuclear dependences for these relative yields.
From this experiment, the ratios
$B\sigma(\Upsilon (2S))/B\sigma(\Upsilon (1S))$ and
$B\sigma(\Upsilon (3S))/B\sigma(\Upsilon (1S))$ over $0 \le  x_F \le 0.6$
are determined as $0.321 \pm 0.012$ and $0.127 \pm 0.009$ for $p+d$.

\begin{figure}[tb]
\includegraphics*[width=\linewidth]{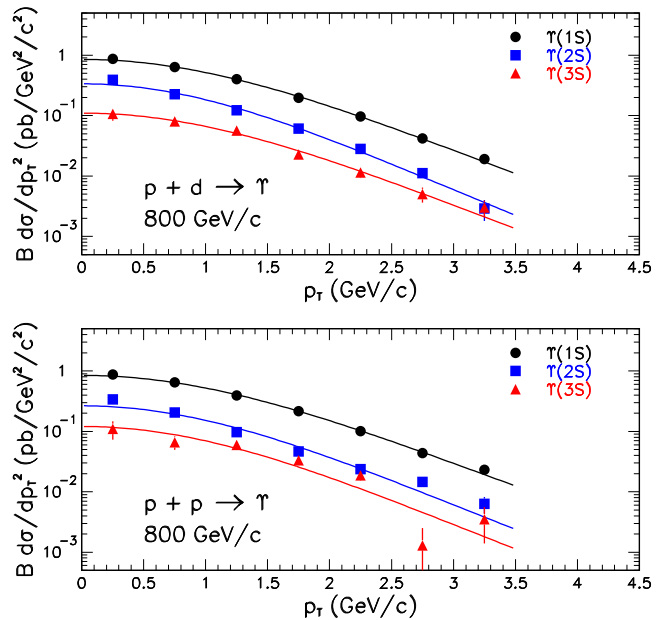}
\caption{(color online). $B\cdot d\sigma/dp_T^2$ (per target nucleon)
for $\Upsilon$ production cross sections for $p+d$ and $p+p$ from the E866
measurement. The curves correspond to fits described 
in the text.}
\label{crosfig3}
\end{figure}

The $p_T$ dependences of the $\Upsilon (1S), \Upsilon (2S)$ and
$\Upsilon (3S)$ cross
sections are listed in Tab.\  I and shown in
Fig.\  3 for $p+d$ and $p+p$. The $d \sigma / d p_T^2$ differential
cross sections are obtained by integrating over the $-1<x_F<1$ range
using a parametrization which best describes the data.
These $p_T$ distributions are fitted with
the parametrization $d\sigma/dp_t^2 = c (1+p_t^2/p_0^2)^{-6}$~\cite{kaplan}
and the results of the fits are shown in Fig.\  3. The values of $p_0$,
listed in Tab.\  I, are somewhat lower than the E605 result~\cite{e605} 
where $p_0 = 3.7$ GeV/c  was obtained.

Figure 4 compares the $\Upsilon (1S)$ production cross section at 800 GeV/c
measured for $p+d$ in E866 and E772~\cite{pat}, and for $p~+$ Cu in
E605~\cite{e605}. The E772 cross sections are roughly a factor of two greater
than those of E866 and E605. Moreover, the shape of the E772 differential
cross sections has a steeper fall-off as $x_F$ increases. To shed some
light on this
apparent discrepancy, calculations for $d\sigma/dx_F$ using the
color-evaporation model (CEM)~\cite{evap} have been performed. 
The CEM
was known to be capable of describing the $x_F$ and energy
dependences of quarkonium production successfully~\cite{bedjidian,vogt}. 
The probability for forming a given quarkonium state is 
treated as a parameter in
this model. As shown in Fig.\  4, the $x_F$ dependences
of both the E866 and the E605 data are well described by calculations
using two different parton distribution functions. In contrast, the 
E772 data show
a steeper $x_F$ dependence than the prediction of the CEM. 
While the origin of the discrepancy between the E866 and the earlier 
E772 results is not understood, the good agreement between the data and the
CEM calculation tends to favor the E866 results.

\begin{figure}[tb]
\includegraphics*[width=\linewidth]{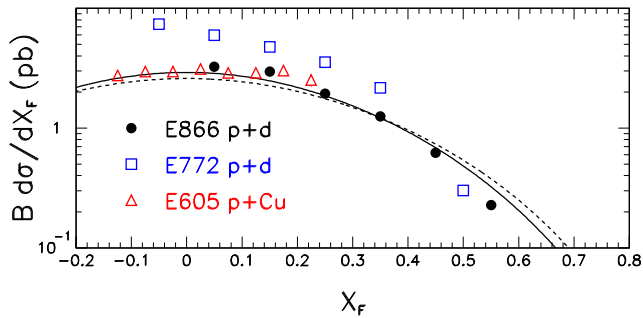}
\caption{(color online). Comparison of the $\Upsilon (1S)$ prodution 
cross sections between E866, E772~\cite{pat}, and E605~\cite{e605}.
The solid and dashed curves correspond to the color-evaporation 
model calculations using the CTEQ4M and CTEQ5L parton distribution 
functions, respectively. The absolute
normalizations of the calculations are adjusted to fit the E866 data.}
\label{crosfig4}
\end{figure}

\begin{figure}[tb]
\includegraphics*[width=\linewidth]{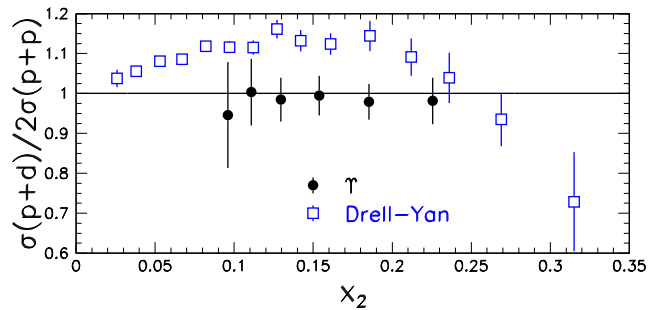}
\caption{(color online). The E866 $\sigma(p + d) /2 \sigma(p + p)$ cross section
ratios for $\Upsilon$ resonances as a function of $x_2$. The
corresponding ratios for the E866 Drell-Yan cross sections~\cite{towell}
are also shown. The error bars are statistical only.}
\label{crosfig5}
\end{figure}

The $\sigma (p+d)/2\sigma(p+p)$ ratios for $\Upsilon (1S+2S+3S)$ production
are shown in Fig.\  5 as a function of $x_2$, the Bjorken-$x$ of the
target parton. Most of the systematic errors cancel for these
ratios, with a remaining $\approx$ 1\% error from the rate dependence
and target compositions~\cite{towell}.
Figure 5 shows that these ratios are consistent with unity, in striking
contrast to the corresponding values~\cite{towell} for the Drell-Yan process,
also shown in Fig.\  5. 
The difference between the Drell-Yan and the
$\Upsilon$ cross section ratios clearly reflect the different
underlying mechanisms in these two processes. The Drell-Yan process,
dominated by the $u - \bar u$ annihilation subprocess, leads to the
relation $\sigma (p + d)_{DY}/2 \sigma (p+p)_{DY} \approx \frac {1}{2} 
(1 + \bar u_n(x_2)/\bar u_p (x_2)) = \frac {1}{2} 
(1 + \bar d_p(x_2)/\bar u_p(x_2))$, where $\bar q_{p,n}$ refers to 
the $\bar q$ distribution in the proton and neutron, respectively. 
For $\Upsilon$ production, the dominance of the
gluon-gluon fusion subprocess at this beam energy
implies that $\sigma (p+d \to \Upsilon)/ 2 \sigma(p+p \to \Upsilon)
\approx \frac {1}{2} (1+g_n(x_2)/g_p(x_2))$. Figure 5 shows that
the gluon distributions in the proton ($g_p$) and neutron ($g_n$) 
are very similar over the $x_2$ range $0.09 < x_2 < 0.25$. The overall
$\sigma (p+d \to \Upsilon)/ 2 \sigma(p+p \to \Upsilon)$ ratio, integrated 
over the measured kinematic range, is $0.984 \pm 0.026 (\rm{stat.}) 
\pm 0.01 (\rm{syst.})$. These results are consistent with no 
charge symmetry breaking effect in the gluon distributions.

In summary, we report the measurement of $\Upsilon$ production 
for $p+p$ and $p+d$ interactions at 800 GeV/c. This measurement 
allows a first determination
of the $\sigma (p+d \to \Upsilon) / 2 \sigma (p+p \to \Upsilon)$ ratio,
which complements the previous measurement of the corresponding Drell-Yan 
ratio. The $\Upsilon$ data indicate that the gluon distributions in the
proton and neutron are very similar. A comparison of the $p+d$ 
data with the previous E605 $p~+$ Cu data shows 
no significant nuclear effects for $\Upsilon$ production in the 
kinematic region near $x_F \sim 0$, consistent with the previous
E772 nuclear-dependence measurement.

This work was supported in part by the U.S. Department of Energy and the
National Science Foundation.


\end{document}